\documentclass[3p, times]{elsarticle}

\setcitestyle{authoryear,open={(},close={)}}

\usepackage{amsmath}
\usepackage{txfonts}
\usepackage{hyperref}
\usepackage[boxed,ruled]{algorithm2e}
\usepackage{url}
\usepackage{graphicx}
\usepackage{geometry}
\usepackage{natbib}

\newcommand{\bs}[1]{\boldsymbol{#1}}

\newcommand{\defeq}{\overset{\underset{\mathrm{def}}{}}{=}}
\title{A simple and fast algorithm for computing discrete Voronoi, Johnson-Mehl or Laguerre diagrams of points}

\author[CNRS]{H.~Moulinec}
\address[CNRS]{
  {Aix-Marseille Univ, CNRS, Centrale Marseille, LMA, 4 Impasse Nikola Tesla,
    CS 40006, 13453 Marseille Cedex 13, France}
}
\begin{document}
\begin{frontmatter}
  \begin{abstract}
    
  This article presents an algorithm to compute \emph{digital images} of Voronoi, Johnson-Mehl or Laguerre
  diagrams of a set of punctual sites, in a domain of a Euclidean space of any dimension.
  The principle of the algorithm is, in a first step, to investigate the voxels in balls centred around
  the sites, and, in a second step, to process the voxels remaining outside the balls. The optimal choice of
  ball radii can be determined analytically or numerically, which allows a performance of the algorithm
  in $O(N_v \ln{N_s})$, where $N_v$ is the total number of voxels of the domain and $N_s$ the number of
  sites of the tessellation.
  Periodic and non-periodic boundary conditions are considered.
  
  A major advantage of the algorithm is its simplicity which makes it very easy to implement.

  This makes the algorithm suitable for creating high resolution images of microstructures containing a
  large number of cells, in particular
  when calculations using FFT-based homogenisation methods are then to be applied to the simulated materials.
\end{abstract}
\end{frontmatter}
\section{Introduction}

\subsection{Generation of microstructures in computational homogenisation}
Over the last three decades, considerable progress has been made in what might be called ``computational homogenisation'',
as the part of computational mechanics dealing with homogenisation or micromechanics.

Until the early 1990s, the available hardware and software computer resources hardly allowed numerical simulations on
realistic microstructures (whether real or simulated) of heterogeneus materials whose geometries are often
very complex.  Therefore, most geometries of the microstructures studied were still very simplified.
To overcome these limitations, the efforts of numerical engineers in the 1990s and 2000s focused on improving existing
methods such as finite elements, for example with the development of multilevel finite element method (\cite{Feyel_2003}),
or developing new, more efficient or more adapted methods, notably the FFT-based homogenisation method introduced by
\cite{Moulinec_1994}.  On the other hand, the progress of computer hardware, in particular the development of new
parallel computers and the software tools (languages and application programming interfaces) that
accompanied them, allowed numerical engineers to improve their codes, often at the cost of adapting numerical methods to
this new paradigm (for example with domain decomposition methods (\cite{Le_Tallec_1991}, \cite{Farhat_Roux_1991}, \cite{Gosselet_2006})).

In parallel with research on calculation methods, studies have been carried out on the microstructures themselves.  The
question of the representativeness of the elementary volume studied was raised with perhaps greater acuity than before
(\cite{Gusev_1997}, \cite{Kanit_2003}).
More recently, since the 2010s, the issue of the generation of artificial microstructures has received more
particular attention. One of the questions is how to create an artificial microstructure that "mimics" a real
microstructure,
(\cite{Fritzen_2009}, \cite{Quey_2011}, \cite{Quey_CMAME_2018}, \cite{Vannuland_2021}).
Another important issue is the efficiency of the algorithms generating the microstructures, whose
performance can differ greatly from one method to another.
The reader is referred to the review article by \cite{Bargmann_2018} which covers the topic of microstructure generation.

\subsection{Context of the study}
\label{context}
The present paper focuses on the generation of Voronoi, Laguerre and Johnson-Mehl diagrams - also called Voronoi, Johnson-Mehl and Laguerre tessellations - which are widely used for the simulation of microstructures of polycrystalline materials.
More precisely, the objective being to calculate the mechanical
behaviour of materials using FFT-based homogenisation methods (\cite{Moulinec_1994}, \cite{Moulinec_1998}), the
motivation of this study is to generate \emph{digital images} of microstructures, i.e. discretisations of
microstructures according to a regular grid, whose elementary data are generally called pixels in the two-dimensional
case and voxels in the three-dimensional case. In the rest of this article, the term voxel is used
for any dimension (2, 3 or higher).

Instead of the term ``digital image'', which is widely used in the image processing community (\cite{Pratt_2007_ch4}), some authors prefer to speak of ``raster images'' as opposed to ``vector images'' or ``vector
graphics'' for which the image is described by geometric shapes.  The object of this study is thus referred to in the
literature as ``raster Voronoi tessellations'' (\cite{Lee_2011}) or ``raster-based method'' (\cite{Chen_1999}),
but more often as ``discrete Voronoi diagrams'' (\cite{Velic_2009}).

A special attention is given, in the present paper, to the case of large images with a high number of grains.
Indeed, it is not uncommon nowadays to apply FFT-based homogenisation methods on microstructures discretised with hundreds
of millions of voxels (\cite{Muller_2015}, \cite{Boittin_2017}, \cite{Chen_2019}, \cite{Vincent_2020},
\cite{Marano_JMPS_2021}
) and, in the case of polycrystalline materials, composed of hundreds of thousands of grains
(\cite{Wojtacki_2020}).  The generation of such microstructures can lead to non negligible computation times. For example, to generate an image of $1000 \times 1000 \times 1000$ voxels of a Voronoi tessellation containing $1 \ 000 \ 000$ grains, the general code Neper of polycrystal generation (\cite{Neper}) requires about $11 \ 300$ seconds on a 8-core processor Xeon Silver 4208 (2.10 GHz).

Moreover, since FFT-based methods make the implicit assumption that the fields involved are periodic,
we are interested, although not exclusively, in the case where the tessellations exhibit periodic conditions at the
boundaries of the spatial domain studied.

An additional constraint imposed by micromechanical studies is that the "sites" or "seeds" defining the tessellation
are not necessarily placed on the exact positions of voxels of the image to be
created. Indeed, the grid of the image must be able to be chosen independently of the layout of the sites, for example
in the case where several realisations of the same tessellation are desired with images of different spatial resolutions.

\subsection{Algorithms for creating tessellations}
\label{section_algorithms_for_creating_tessellations}
Simply put, a Voronoi tessellation of a given set of discrete points in a Euclidean space, hereafter referred to as sites,
is a tiling of the space into regions, each of which gathers all points in the space closer to a given site
than to any other.
Laguerre and Jonhson-Mehl tessellations can be considered as variants of Voronoi tessellations, where the function
evaluating the proximity of one point to another is different from the Euclidean distance.
More precise definitions are given in section \ref{section definitions}.

Since their introduction (\cite{Voronoi_1908}), Voronoi tessellations gave rise to an abundant literature.  Many authors
have been interested in the mathematical properties of these constructions (see \cite{Stoyan_1995},
\cite{Aurenhammer_1991} or \cite{Okabe_2000} for Voronoi tessellations, \cite{Moller_1992} for those of Johnson-Mehl and
\cite{Aurenhammer_1991}, \cite{Okabe_2000} or \cite{Lautensack_Zuyev_2008} for Laguerre tessellations) and, in particular, in how to describe and
calculate the cells of the tiling of the space that constitutes them.  Numerous algorithms have been proposed, among
which two families can be distinguished schematically.

\begin{enumerate}
\item A first category of algorithms, mainly from the robotics and computer vision communities, are interested in the
calculation of the Euclidean Distance Transform (EDT), which consists in evaluating for each voxel of an image its
distance to a given region of interest of the image (see among others \cite{Saito_1994}, \cite{Breu_1995},
\cite{Hirata_1996}).  When these regions of interest are composed of isolated voxels, this problem reduces 
to that of creating an image of a Voronoi tessellation. See \cite{Fabbri_2008} for a review and comparison
of existing algorithms.
Several of the proposed algorithms achieve the optimal performance in $O(N_v)$, where $N_v$ is the total number of
voxels in the image.

\item Another family of methods, mainly from the computational geometry community, determines the geometric parameters of
the cells (for example, the vertices, edges, facets of the polygons or polyhedra which constitute the cells of the
Voronoi tessellations in the two-dimensional or three-dimensional cases, respectively). An extensive review of the
existing algorithms for Voronoi diagrams, and a comparison of their efficiency can be found in \cite{Okabe_2000}. It
appears that some algorithms attain the optimal time complexity $O(N_s)$, where $N_s$ is the number of sites, but that
many methods behave in $O(N_s \ln N_s)$, which is the ``worst-case optimal time complexity'', or even worse.
\end{enumerate}

However, these algorithms have a number of drawbacks for the problem addressed in this paper.
Although the approximations on the calculation of the distance in early EDT algorithms, imposed largely by the technical
limitations of the time, have been overcome today with algorithms that achieve what is called ``Exact EDT'', all
of the proposed algorithms, to the best of the author's knowledge, place the sites of the Voronoi cells in
the exact location of a voxel, which is inadequate for the present study.

For the case of interest of this study, i.e. the creation of images of tessellations, the algorithms of the second type must
be completed by a step where it must be determined to which cell of the tessellation each voxels of the image belongs.
So to speak, after having described the geometry of the cells of the tessellation, the algorithm must ``fill''
the cells with the voxels they contain.  This option is rarely implemented in the existing codes but is proposed by the
Neper code  (\cite{Quey_2011}, \cite{Neper}), where it is called ``raster tessellation meshing''.
As the whole numerical methods in
question are rather delicate to implement, materials science researchers often opt either to use Neper or to implement a
simple brute force algorithm
whose low performance
in $O(N_v \, N_s)$ makes it unsuitable for large simulations.

The aim of this article is to propose a simple algorithm, just a little more difficult to implement than the brute force algorithm,
but with much better performance in $O\big(N_v \, \ln N_s \big)$.

First,  in section \ref{section definitions} of the article, the tessellations of Voronoi, Johnson-Mehl and Laguerre
are briefly introduced.
In section \ref{section algorithms}, an accelerated algorithm is presented for the different types of tessellations and
its performance is compared to that of Neper code, in section \ref{section comparison}.

\section{Voronoi, Johnson-Mehl and Laguerre tessellations}
\label{section definitions}

\subsection{Voronoi tessellations}
A Voronoi tessellation, also called Voronoi diagram,  associated with a given set ${\cal S}=\{\bs{x}_s\}_{s=1,...,N_s}$
of $N_s$ points, called ``sites'', ``seeds'' or ``generators'',
of a given domain $\Omega$ of a Euclidean space,
is a division of the domain into $N_s$ zones, called ``cells'',
each cell $s$ including all the points
of the domain that are closer to the site $s$ of position $\bs{x}_s$ than to any other site in the set.
Thus, the cell $C_s$ associated to the site $s$ of the set ${\cal S}$ is a subdivision of domain $\Omega$ defined by
\begin{equation}
  C_s = \{
  \bs{x} \in {\Omega} \ | \
    d(\bs{x},\bs{x_s}) \le d(\bs{x}, \bs{x_{s'}}), \forall s' \in {\cal S}
    \}
    \ ,
    \label{Voronoi_Cell}
\end{equation}
where $d(\bs{x},\bs{y})$ denotes the distance between the two points $\bs{x}$ and $\bs{y}$.
Throughout the article, multidimensional quantities, such as vectors $\bs{x}$ and $\bs{x_s}$ in
relation \ref{Voronoi_Cell}, are noted in bold type while scalar data are in normal type.

The distance $d(,)$ used above can be the usual Euclidean distance,
but in the case of periodic conditions on the boundaries of domain $\Omega$,
the Euclidean distance can advantageously be replaced by what is called hereafter the
L-periodic distance. The reader is referred to \ref{L-distance appendix} for the definition (\ref{L-distance_def}) and calculation (\ref{dL}) of the 
L-periodic distance.  In the rest of the article, the distance $d(,)$ denotes either the Euclidean distance or
the L-periodic distance, depending on the boundary conditions considered.

In practical applications, the Euclidean space considered is most often two- or three-dimensional,
but the definition of the Voronoi tessellation given above is valid for any finite dimension.

\begin{figure}[htp!]
\begin{center}
\includegraphics[width=12cm]{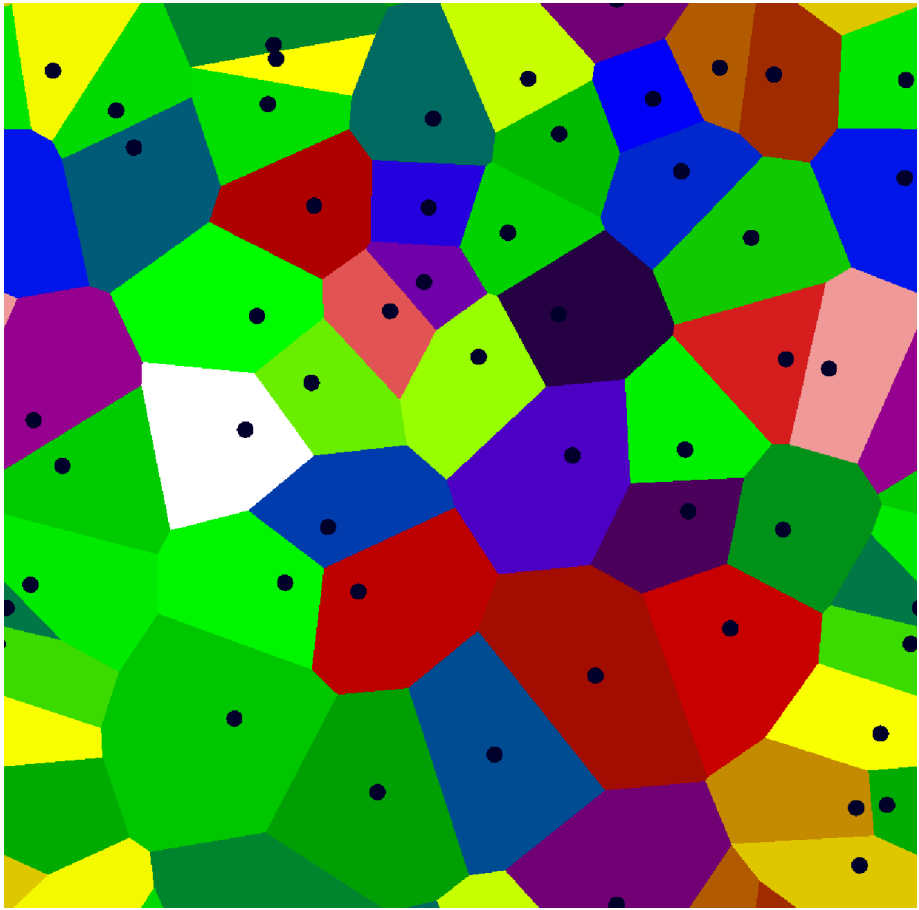}
\end{center}
\caption{
  Voronoi tessellation of 50 punctual sites, with periodic conditions.
  The cells are plotted in arbitrary colours.
  The sites are marked by small black circles.
}
\label{voronoi01}
\end{figure}

An example of a two-dimensional Voronoi tessellation, generated from
50 sites randomly arranged in the area,  is shown in Figure \ref{voronoi01}.
Periodic conditions on the borders of the domain have been assumed.

\subsection{Johnson-Mehl tessellations}
The model of the Johnson-Mehl tessellation, also called Johnson-Mehl-Avrami-Kolmogorov tessellation,
has been developed to account for crystallisation in a material (\cite{ Johnson_Mehl_1939}).
To put it simply, a crystal grows from a site that appears in the volume at a given location
and at a given {time}, until its boundaries reach those of other crystals.
The tessellation is the resulting division of the space,
when all the crystals have finished growing and have therefore completely covered the domain of space
$\Omega$ considered.

Thus, considering the set of sites given by
\begin{equation}
  {\cal S} = \{ (\bs{x_s}, t_s) \}_{s=1,2,...,N_s}
  \ ,
\end{equation}
where $\bs{x_s}$ is the position of site $s$ and $t_s$ is its birth time
(i.e. the time at which the site appears in ${\Omega}$ and starts to grow),
a given point $\bs{x}$ of the space would be reached by the growth of the crystal generated by
site $s$ - in the absence of interactions with any other site - at time $t$ which is
\begin{equation}
  t = d(\bs{x}, \bs{x_s}) / {G} + t_s
  \ ,
  \label{JMAK2}
\end{equation}
where $G$ is the radial growth of the crystals, supposed to be constant and to be identical for all the crystals.

Thus, the cell associated with a given site $s$ is defined by
\begin{equation}
  C_s = \{
  \bs{x} \in {\Omega} \ | \
    d(\bs{x},\bs{x_s})/G + t_s \le d(\bs{x}, \bs{x_{s'}})/G + t_{s'}, \forall s' \in {\cal S}
    \}
    \label{JM1}
\end{equation}
or, equivalently,
\begin{equation}
  C_s = \{
  \bs{x} \in {\Omega} \ | \
    d(\bs{x},\bs{x_s}) + G t_s \le d(\bs{x}, \bs{x_{s'}})+ G t_{s'}, \forall s' \in {\cal S}
    \}
    \ ,
    \label{JM2}
\end{equation}
$G$ being supposed to be positive.

\subsection{Laguerre tessellations}
\label{Laguerre_section}

A Laguerre tessellation, also called power diagram,  is generally presented in the literature as a partition
of the space in cells defined by a set of spheres $s$ of centres $\bs{x}_s$ and radii $r_s$.
A given point $\bs{x}$ belongs to the cell $C_s$ associated with the site - or sphere - $s$ which minimises
the so-called ``power distance'' $p_d(\bs{x},s)$
defined by
\begin{equation}
  p_d(\bs{x},s) = d^2(\bs{x},\bs{x_s}) -r_s^2 .
\end{equation}
Hence, the cell associated to the site $s$ is defined by
\begin{equation}
  C_s = \{
  \bs{x} \in {\Omega} \ | \
  p_d(\bs{x},s)
  \le
  p_d(\bs{x},{s'}) , \forall s' \in {\cal S}
    \}
\end{equation}

Alternatively, from a physical point of view and as detailed in \ref{Laguerre_appendix}, a Laguerre tessellation
can be considered as the resulting partition of a  crystallisation process starting from a set of sites
${\cal S} = \{ (\bs{x_s}, t_s) \}_{s=1,2,...,N_s}$, 
where $\bs{x}_s$ is the location of the site $s$
and $t_s$ is its birth time,
with a constant growth ${G}$ of the square of the radii of the crystals.
The principle remains similar to that of Johnson-Mehl tessellation: crystal $s$ reaches a given point
$\bs{x}$ at time $t$ given by
\begin{equation}
  t = d^2(\bs{x}, \bs{x}_s) / G + t_s \ ,
  \label{LAGUERRE1}
\end{equation}
and finally the cell $C_s$ associated to a given site $s$ is defined by
\begin{equation}
  C_s = \{
  \bs{x} \in {\Omega} \ | \
    d^2(\bs{x},\bs{x_s}) + G t_s \le d^2(\bs{x}, \bs{x_{s'}})+ G t_{s'}, \forall s' \in {\cal S}
    \}
    \ .
    \label{LAGUERRE2}
\end{equation}
This notation has the advantage of making the definitions of Johnson-Mehl and
Laguerre tessellations similar.

\paragraph{Remark}
In the particular case where all sites appear at the same time,
or in the case $G$ is zero, when using definitions \ref{JM2} and \ref{LAGUERRE2},
Johnson-Mehl and Laguerre tessellations reduce to a Voronoi tessellation.

\section{Tessellation algorithms}
\label{section algorithms}
In this section, algorithms are presented that solve the problem studied, the terms of which are briefly recalled here.
\begin{itemize}
\item The algorithms must create a \emph{digital image} of a Voronoi, Johnson-Mehl
  or Laguerre tessellation  of a set of punctual sites.
\item The positions of the sites are arbitrary and do not necessarily coincide with voxel centres.
\item The distance used to estimate the proximity of a voxel to a site must be \emph{exact}.
\item The boundary conditions of the image created can be periodic
  or non-periodic.
\item Although most applications of Voronoi tessellation relate to two- or three-dimensional spaces, the algorithms
  should apply to spaces of any dimension.
  
\item The domain $\Omega$, in which the tessellation is created, is a d-dimensional rectangular parallelepiped,
  i.e. a volume of the Euclidean space of dimension $d$, defined by $\Omega=[0,L_1) \times ... \times [0,L_d)$
  in a given orthonormal basis.
  
\end{itemize}

In addition, the arrangement of the sites in the volume are considered purely random,
following a uniform probability distribution.
In the case of the Voronoi diagram, this type of tessellations is sometimes called "Poisson-Voronoi tessellations".

The computational cost of the algorithms are estimated by counting the total number of calculations of distance 
between points performed during the execution of the algorithms (and in some cases, the number of calculations of the square of the distance).
The effect of memory access and compilation optimisation are not considered, as it largely depends on
the system on which the algorithms are implemented.

\subsection{A brute force algorithm}
\label{brute force algo}

The most simple algorithm to build an image of a Voronoi tessellation consists of calculating for each
voxel its distance to all the sites and thus determining the closest site.
Diagram \ref{simple Voronoi algorithm} schematises this algorithm.
Similar algorithms can easily be devised to Johnson-Mehl or Laguerre tessellations.

 \begin{algorithm}
   \SetAlgoLined
   \DontPrintSemicolon
   \SetKwInOut{Input}{input}\SetKwInOut{Output}{output}

   \Input{a set of $N_s$ sites of positions $\{\bs{x}_s\}$ in the unit cell
     ${\Omega}=[0,L_1) \ \times \ [0,L_2) \ \times \ ... \ \times \ [0,L_d)$
               }

    \Output{
      \begin{itemize}
      \item an image ${\cal I}$ of $\bs{n}=(n_1,n_2,...,n_d)$ voxels describing the tessellation
        in space domain ${\Omega}$
      \item  an image ${\cal D}$ of $\bs{n}=(n_1,n_2,...,n_d)$ voxels giving the distance
        of each point $\bs{x(k)}$ to the closest Voronoi site
      \end{itemize}
    }
   \BlankLine
   \ForAll{$\bs{k} \in \{1,...,n_1\}\times\{1,...,n_2\}\times ... \times \{1,...,n_d\}$}{
     ${\cal D}(\bs{k})=\infty$ \;
     \ForAll{$s = 1$ to $N_s$}{
       \If{ $d({\bs{x(k)},\bs{x}_s}) < {\cal D}({\bs{k}})$ }{
         ${\cal I}({\bs{k}})=s$ \;
         ${\cal D}({\bs{k}})= d({\bs{x(k)},\bs{x}_s})$ \;
       }
     }
   }
   \caption{Brute force algorithm to create an image of a Voronoi tessellation from a set of
     positions of sites in a Euclidean space of dimension $d$.
     $\bs{k}$ gathers the $d$ indices of the voxels in the image. $\bs{x(k)}$ is the location of voxel $\bs{k}$.
   }
   \label{simple Voronoi algorithm}
 \end{algorithm}
In the case of Voronoi and Laguerre tessellations (but not of Johnson-Mehl tessellations), a substantial
gain in efficiency can be achieved by considering the square of the distance
when comparing the distances between a given voxel and all the sites, as this saves the costly calculation of a square
root repeated many times.  Despite this improvement, the method still performs poorly, the total cost of the
algorithm being in $O( N_v N_s )$, where $N_v$ is the total number of voxels in the image and
$N_s$ is the number of sites.

\subsection{An accelerated Voronoi tessellation algorithm}
\label{accelerated algo}

\subsubsection{Principle of the algorithm}
Starting from the observation that the cells of a Poisson-Voronoi tessellation are all fairly similar in size,
the principle of the algorithm consists first in examining a given neighborhood of
all the sites, and determining for each voxel in the neighborhood whether the site considered is the closest,
and then, in a second step, in processing the remaining voxels with the ``brute force'' algorithm described above.

The algorithm is detailed below.

\subsubsection{Detail of the algorithm}
We start by choosing balls of investigation, ${\cal B}_s$, centred on each site $\bs{x}_s$ of $\cal S$
with a given radius $r_0$ and a corresponding volume $v_0$:
\begin{equation}
  {\cal B}_s = \{ \bs{x} \in \Omega | \ d(\bs{x},\bs{x_s}) \le r_0 \} \ .
\end{equation}
The choice of $r_0$ and $v_0$ is discussed in section \ref{section_optimal_choice_Voronoi}.
The voxels inside each investigation ball
are considered and their distance to the associated site is evaluated and saved.

After all the balls have been examined, two cases are possible for a given voxel
of domain $\Omega$:
\begin{itemize}
\item It belongs to one or more balls. In other words, the distance between the voxel and all the sites is
  smaller than ${r_0}$ only for the sites that are the centres of these balls.
  Thus the voxel belongs to the cell of the nearest of these sites.
\item It belongs to no balls.
\end{itemize}

To complete the tessellation, the distance between each voxel of  the second case
and all the sites are evaluated and the voxel is assigned to the cell of the closest site.

Diagram \ref{faster_voronoi_algo} summarises the algorithm that performs these tasks.

\begin{algorithm}
  \SetAlgoLined
  \DontPrintSemicolon
  \SetKwInOut{Input}{input}\SetKwInOut{Output}{output}
  
  \Input{
    \begin{itemize}
    \item a set of $N_s$ sites of positions $\{\bs{x}_s\}$ in space domain
      ${\Omega}=[0,L_1) \ \times \ [0,L_2) \ \times \ ... \ \times \ [0,L_d)$
    \end{itemize}
    }
  \Output{
    \begin{itemize}
      \item {an image ${\cal I}$ of $\bs{n}=(n_1,n_2,...,n_d)$ voxels describing the Voronoi tessellation
        in space domain $\Omega$}
      \item  an image ${\cal D}$ of $\bs{n}=(n_1,n_2,...,n_d)$ voxels giving the distance
        of each point $\bs{x(k)}$ to the closest Voronoi site
        (i.e. the map of the Euclidean Distance Transform)
    \end{itemize}
  }
  \BlankLine
  Initialisation: \;
  radius $r_0$ of the balls of investigation is calculated using relation \ref{v0_3d} \;  
  \ForAll{$\bs{k} \in \{1,...,n_1\}\times\{1,...,n_2\}\times ... \times \{1,...,n_d\}$}{
    ${\cal D}(\bs{k}) = \infty$ \;
  }
  \BlankLine
  Step 1: \;
  \ForAll{ $s = 1, 2, ..., N_s$ }{
        \ForAll{ $\bs{k} \ | \ \bs{x(k)} \in {\cal B}_s (\text{Ball of centre } \bs{x}_s \ \text{and  radius } {r_0}) $} {
        \If{ $d(\bs{x(k)},\bs{x_s}) < {\cal D}(\bs{k})$ }
           {
             ${\cal I}(\bs{k}) =  s$ \;
             ${\cal D}(\bs{k}) =  d(\bs{x(k)},\bs{x_s})$ \;
           }
    }         
  }
  \BlankLine
  Step 2: \;
  \ForAll{$\bs{k} \in \{1,...,n_1\}\times\{1,...,n_2\}\times ... \times \{1,...,n_d\}$}{
    \If{${\cal D}(\bs{k}) = \infty$}{
      \ForAll{$s = 1, 2, ..., N_s$}{
        \If{ $d(\bs{x(k)},\bs{x_s}) < {\cal D}(\bs{k})$}{
          ${\cal I}(\bs{k}) =  s$ \;
          ${\cal D}(\bs{k}) =  d(\bs{x(k)},\bs{x_s})$ \;
        }
      }
    }
  }
  \caption{Faster algorithm to create an image of a Voronoi tessellation
    in a Euclidean space of dimension~$d$}
  \label{faster_voronoi_algo}
\end{algorithm}

\paragraph{Remarks}

Similarly to what has already been mentioned in section \ref{brute force algo}, the algorithm can be accelerated by considering the square of the distance instead
of the distance.

A by-product of Algorithm \ref{faster_voronoi_algo} is the image $\cal D$ of the EDT
(see section \ref{section_algorithms_for_creating_tessellations}).

\subsubsection{Optimal choice of the size of the investigation ball}
\label{section_optimal_choice_Voronoi}
The choice of the radius $r_0$ of the investigation balls plays a crucial role on the efficiency of
Algorithm  \ref{faster_voronoi_algo}.
When $r_0$ is large, the number of voxels inside the balls and thus the number of calculations of
their distance to the centres of the balls are high. Inversely, the number of voxels outside the
balls is small and becomes zero when the balls cover the whole domain $\Omega$.
When $r_0$ is small, the situation is the opposite: the number of distance calculations in the first step
of the algorithm is small and the one of the second step is high.
Between these two extremes, there is an optimal choice for the size of the investigation balls.

The number $n_{step~1}$ of distance calculations in the first step of Algorithm \ref{faster_voronoi_algo}
is equal to the total number of voxels inside the investigation balls.
Neglecting the slight fluctuations of the number of voxels from one ball to another,
due to spatial discretisation, one has
\begin{equation}
  n_{step~1} = N_v   \frac{ \sum_s v'_s}{V}
  \ ,
  \label{nstep1}
\end{equation}
where $V$ is the volume of $\Omega$ and
where $v_s'$ is the volume of intersection of the ball of investigation ${\cal B}_s$ with the domain of interest, that is
\begin{itemize}
\item in the non-periodic case: $v'_s$ is the volume of the intersection of the ball ${\cal B}_s$ and the
  domain $\Omega$.
  The balls of investigation crossing the boundaries of $\Omega$ must be truncated from their part outside $\Omega$,
  resulting in a reduction of the number of voxels to be investigated; in that case, one has $v_s'<v_0$. 
  In the case when the ball is entirely inside the domain $\Omega$, one has $v'_s = v_0$.
  And in the extreme case when the ball encompasses the whole domain $\Omega$, one has $v'_s=V$.
\item in the periodic case: $v'_s$ is the volume of the intersection of the ball ${\cal B}_s$ and the
  domain $\Omega_s$ which is the translation of parallelepiped $\Omega$, centred on $\bs{x}_s$.  As long as
  the ball is smaller than the sphere inscribed in $\Omega_s$ (i.e. $r_0 \le L_i/2, \forall i=1,...,d$), one has $v'_s =
  v_0$. When the ball is bigger than the circumscribed sphere (i.e. $r_0 \ge \frac{1}{2}\sqrt{\sum_{i=1}^d L_i^2}$), one
  has $v'_s=V$. And between these two cases, $v'_s$ takes intermediate values.
\end{itemize}
With Voronoi tessellations, the conditions ensuring that $v'_s=v_0$ (small balls compared to the domain size)
are satisfied in most of the cases, and, in the periodic case, relation \ref{nstep1} becomes
\begin{equation}
    n_{step~1}
    =  N_v   N_s v_0  \ .
    \label{nstep1b}
\end{equation}
If non-periodic conditions are assumed, one has only an upper bound,
\begin{equation}
    n_{step~1}
    \le  N_v   N_s v_0
    \ ,
    \label{nstep1c}
\end{equation}
which tends to be exact when the number of grains increases and thus when the ratio of spheres crossing the boundaries
of $\Omega$ decreases.  In the following, expression \ref{nstep1b} is used, as a fair approximation of the
algorithm behaviour.

\vspace{0.3cm}
The number $n_{step~2}$ of estimation of the distances in the second step of Algorithm \ref{faster_voronoi_algo}
is given by
\begin{equation*}
    n_{step~2}
    = N_s n_{out}
    \ ,
\end{equation*}
$n_{out}$ being the number of voxels at a distance greater than ${r_0}$ from all sites.
The value of $n_{out}$ depends on the set of sites, nevertheless, its statistics can be determined.
When a site $s$ is randomly placed in the volume, the probability for a given point of the domain to be outside the
ball ${\cal B}_s$ of radius ${r_0}$ centred on the site is given by
\begin{equation}
  p_s = 1-\frac{ {v'_s} }{V} \ .
  \label{p1a}
\end{equation}
In the same manner as above, $v'_s$ is approximated by  $v'_s=v_0$ and thus
\begin{equation}
  p_s = 1-\frac{ {v_0} }{V} \ .
  \label{p1}
\end{equation}
As the positions of the $N_s$ sites are supposed to be uncorrelated, the probability for a given point to remain
outside the $N_s$ balls of volume ${v_0}$ centred on the sites, is
\begin{equation*}
  p_{N_s} = \Pi_{s} p_s \ ,
\end{equation*}
where the notation $\Pi_s$ means the product for all $s\in {\cal S}$. Hence
\begin{equation*}
  p_{N_s} = \big( 1-\frac{v_0}{V} \big)^{N_s}\ .
\end{equation*}
Finally, the mean number of voxels outside all the balls is given by
\begin{equation*}
  \text{E}(n_{out}) =  N_v \big( 1-\frac{v_0}{V} \big)^{N_s} 
\end{equation*}
and the mean number of distance calculations in step 2 is
\begin{equation*}
  \text{E}(n_{step\ 2}) =  N_s N_v \big( 1-\frac{v_0}{V} \big)^{N_s}
  \ .
\end{equation*}
Thus, the total number of distance calculations in Algorithm \ref{faster_voronoi_algo} is estimated by
\begin{equation}
  E(n_{step~1 + \ 2})= N_s N_v \bigg( \frac{v_0}{V} + \big( 1-\frac{v_0}{V} \big)^{N_s} \bigg)
  \ .
  \label{nstep12_3d} 
\end{equation}
The optimal choice for $v_0$, i.e. the value of $v_0$ which minimises $E(n_{step \ 1+2})$,
is therefore easy to calculate:
\begin{equation}
  \frac{v_0}{V} = 1-\bigg( \frac{1}{N_s} \bigg)^{\frac{1}{N_s-1}}
  \ .
  \label{v0_3d}
\end{equation}
With this choice of $v_0$ one has
\begin{equation*}
    E(n_{step~1 + \ 2})
    =
    N_v N_s 
    \Bigg(
    1
    +
    \Big( \frac{1}{N_s} \Big)^{\frac{1}{N_s-1}}
    \Big(
    \frac{1}{N_s}
    - 1
    \Big)
    \Bigg)
    \ .
\end{equation*}
When the number $N_s$ of sites is high,
one has
\begin{equation}
  n_{step~1 + \ 2} =
  N_v \ln N_s + N_v + o(N_v)
  \ .
  \label{dl}
\end{equation}
Thus, the computational performance of the algorithm is in
$O\big ( N_v \ln N_s  \big)$, which 
is significantly more efficient than the one of Algorithm \ref{simple Voronoi algorithm} for which the total
number of distance calculations is in $O(N_v \, N_s)$.

When different configurations are considered for which the number of sites $N_s$ vary but the resolution,
defined as the mean number of voxels per Voronoi cell, remains constant,
the performance of the algorithm at constant resolution
can be considered as being in $O \big( N_s \ln N_s \big)$
while the performance of the brute force algorithm is in $O(N_s^2)$.

\begin{figure}[htp!]

\begin{center}
\includegraphics[angle=-90,width=12cm]{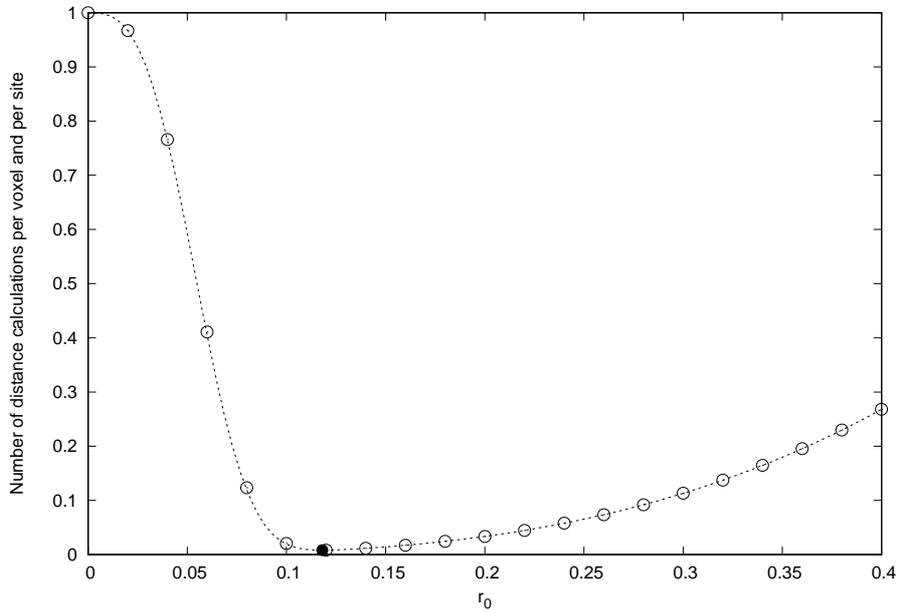}
\end{center}

\caption{
  Generation of an image of $200\times 200 \times 200$ voxels of a Voronoi diagram of 1000 sites
  with Algorithm \ref{faster_voronoi_algo} : 
  number of calculations of distances between voxels and sites per voxel and per site for different
  values of parameter $r_0$ (radius of the balls of investigation around the sites).
  Empty circles: actual number of distance calculations during the execution of the algorithm.
  Dashed line: model given by relation \ref{nstep12_3d}.
  Black circle: optimal radius $r_0 \simeq 0.12$, estimated using relation \ref{v0_3d}.
  }
\label{voro4}
\end{figure}

The number of distance calculations as a function of the radius $r_0$, deduced from relation \ref{nstep12_3d},
is plotted in Figure \ref{voro4} and compared with the actual
numbers of distance calculations made during an execution of the algorithm, applied to
the creation of an image of $200\times 200 \times 200$ voxels of a Voronoi tessellation from $1000$
given sites.
The correspondence is significant and the minimum of the curve is well estimated by relation \ref{v0_3d}.

\begin{figure}[htp!]

\begin{center}
  \includegraphics[angle=-90,width=12cm]{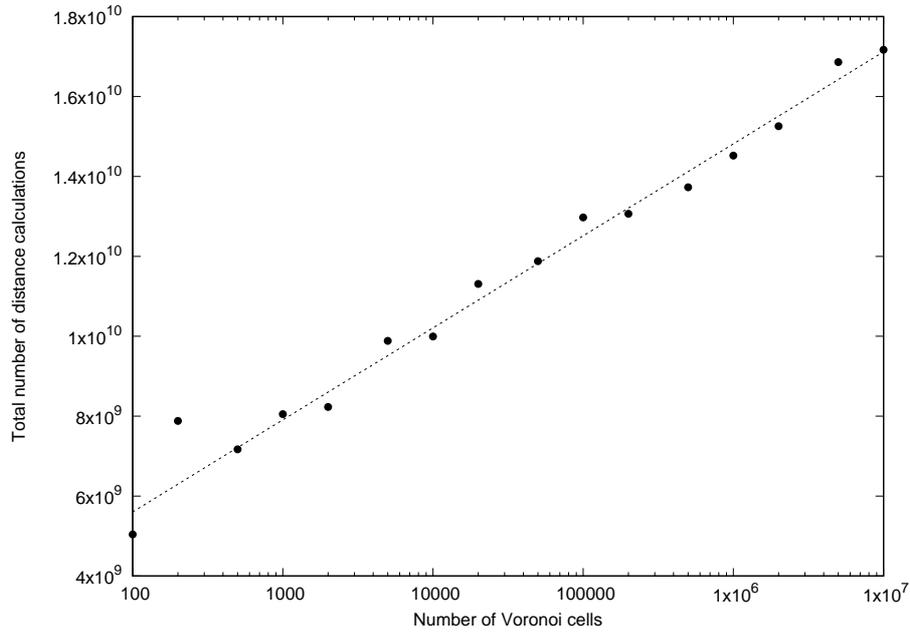}
\end{center}

\caption{ Number of calculations of distances between voxels and sites in the fast algorithm
  in the case of an image of $1000 \times 1000 \times 1000$ voxels
  of a Voronoi tessellation for different numbers of sites $N_s$.
  Measure of the numbers of
  distance calculations actually performed during the execution of the algorithm (circles).
  Expression \ref{dl} approximating the behaviour of the algorithm for large numbers of cells (dashed line).
  }
\label{fdl}
\end{figure}

Expression \ref{dl}, approximating the behaviour of the fast algorithm for large numbers of
sites, is compared in Figure \ref{fdl} with the actual number of distance calculations for the case of a
$1000\times 1000 \times 1000$ images with varying numbers of cells, again leading to an appreciable agreement.

\paragraph{Remark}
When the numbers of sites becomes very large, the expression \ref{v0_3d} of volume $v_0$ can be approximated by
\begin{equation*}
  \frac{v_0}{V} \simeq \frac{\ln{N_s}}{N_s} \ .
\end{equation*}
Comparing it with the average volume of the cells $<v_s>=V/N_s$, one has
\begin{equation*}
  \frac{v_0}{<v_s>} \simeq \ln{N_s} \ .
\end{equation*}
This ratio never stops increasing, which reflects the increasing computational weight of step 2 as $N_s$ grows.
This effect could be reduced by using a more efficient algorithm than the brute force algorithm in step 2.
However, this point is developed in this article, for simplicity.

\begin{figure}[htp!]
\begin{center}
  \includegraphics[width=7.2cm]{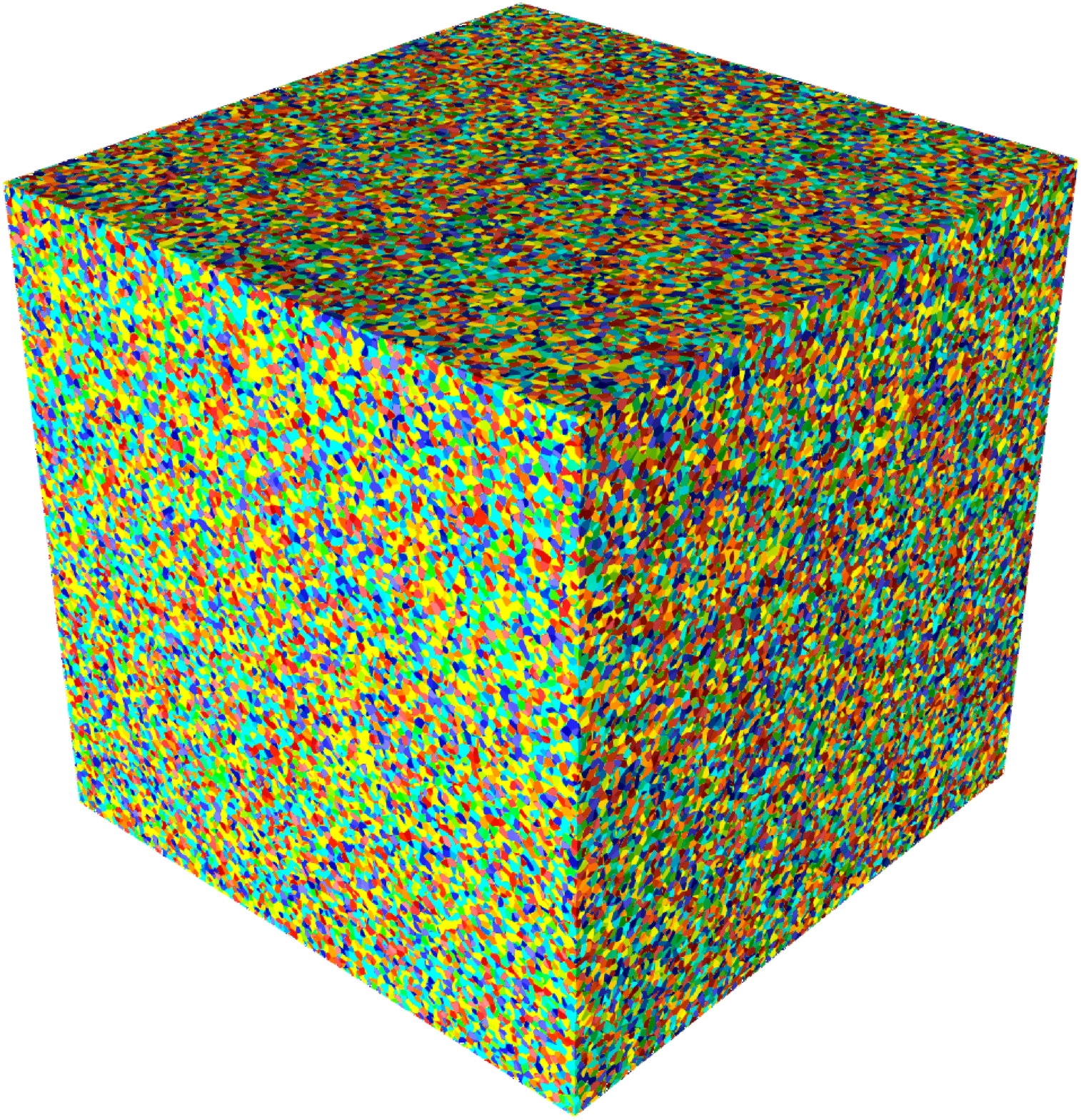} \hspace{0.5cm}
  \includegraphics[width=7.2cm]{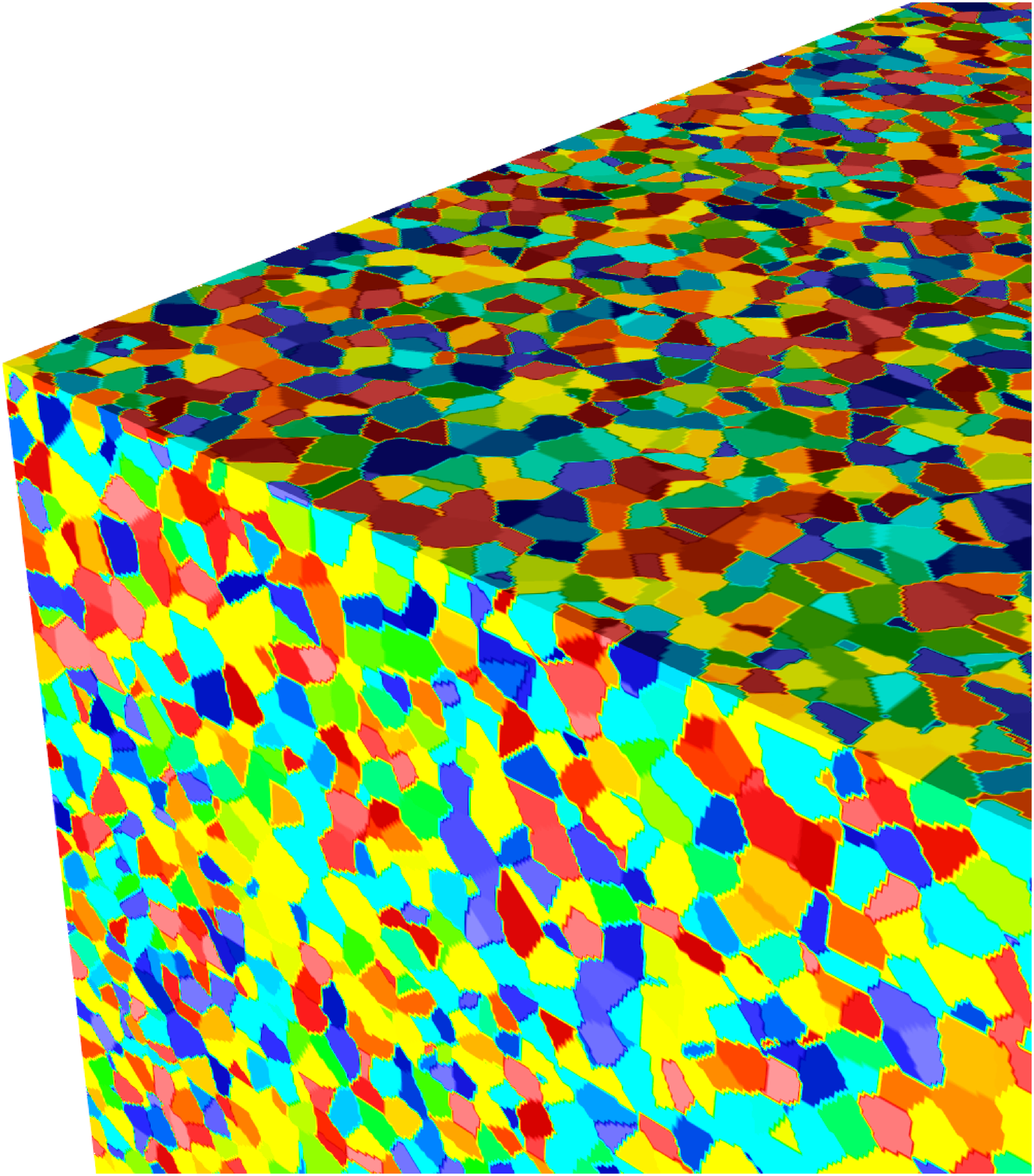}
\end{center}
\caption{
  Image of $1000\times 1000 \times 1000$ voxels of a Voronoi tessellation of $1 \ 000 \ 000$  punctual sites,
  generated by Algorithm \ref{faster_voronoi_algo} in 77.2 seconds on a 8-core processor Xeon Silver 4208
  (2.10 GHz). The cells are plotted in arbitrary pseudo-colours.
  Left: general view of the volume. Right: detail. 
}
\label{voronoi02}
\end{figure}

An example of realisation of microstructure with Algorithm  \ref{faster_voronoi_algo}  is presented in
Figure \ref{voronoi02}.

\subsection{An accelerated Johnson-Mehl tessellation algorithm}
\label {fast_jmak_section}

\subsubsection{Description of the algorithm}
The principle of the accelerated algorithm remains the same for Johnson-Mehl tessellations, except that
the quantity used to determine the site of the cell which a given voxel belongs to, is no longer
the Euclidean distance, but the time $t$ at which the growth of the crystal centred on the site reaches the point,
given by relation \ref{JMAK2}, which can be recalled here:
\begin{equation*}
  t = d(\bs{x}, \bs{x_s}) / {G} + t_s \ .
  \label{JM01}
\end{equation*}

The algorithm for the Johnson-Mehl tessellation thus resumes to the following: at a given time $t_0$, whose choice is discussed in section \ref{section_optimal_choice_Johnson_Mehl}, the spherical region of the fictitious crystallisation
process of each site is examined. The radius $r_s$ of the ball of site $s$ is given by
\begin{equation}
  r_s = G (t_0-t_s)^+
  \ ,
  \label{rs JM}
\end{equation}
where $( . )^+$ denotes the positive part.
The voxels in the ball  of radius $r_s$ centred on the site $x_s$, i.e. the voxels satisfying
\begin{equation}
  d(\bs{x}, \bs{x_s}) \le r_s \ ,
  \label{JM02}
\end{equation}
are examined and the time they are reached by the crystal growth is evaluated with relation \ref{JMAK2}.

After this step, there are two different cases for each voxel of the image:
\begin{itemize}
\item 
  it belongs to one or more balls, i.e. relation \ref{JM02} 
  is satisfied for one or more sites, 
  thus the voxel belongs to the cell of the one of these sites which minimises $d(\bs{x}, \bs{x_s}) / {G} + t_s$,
\item it belongs to no balls. In that case, the value of $d(\bs{x}, \bs{x_s}) / {G} + t_s$
  is calculated for all the sites
  and the voxel belongs to the cell of the site which minimises it.
\end{itemize}

The algorithm is summarised in Diagram \ref{faster_jmak_algo}.
The differences with the algorithm for Voronoi tessellations are that
the radii of the balls around the sites vary according to the sites,
and that the quantity used to evaluate the proximity of a site is different.

\subsubsection{Optimal choice for $t_0$}
\label{section_optimal_choice_Johnson_Mehl}
The value of $t_0$ must be chosen to minimise the computational cost of the algorithm, such as $v_0$ in the algorithm
for Voronoi tessellations.

The cost of the first step of the algorithm is proportional to the number of voxels in the balls
centred on the sites, i.e.
\begin{equation}
  n_{step~1} = N_v   \frac{ \sum_s v'_s}{V}
  \ ,
\end{equation}
where $v'_s$ is defined as in the case of Voronoi tessellation:
it is the volume of the intersection of the ball ${\cal B}_s$ and of the domain $\Omega_s$ ( respectively $\Omega$)
in the periodic case (respectively in the non-periodic case), except that the balls ${\cal B}_s$ have radii
$r_s$ depending on the sites while they have an identical radius $r_0$ in the Voronoi algorithm.

When a ball is smaller than the sphere inscribed in the domain, one has
$$v'_s = v_s \ ,$$
where $v_s$ is the volume of the whole ball ${\cal B}_s$ ( i.e. the volume of a ball of radius $r_s$ of
relation \ref{rs JM}).
When the ball encompasses the whole domain, one has
$$v'_s = V \ .$$
The exact expression of $v'_s$, taking into account the case when the ball is between these two extreme cases
is uselessly complicate and can be approximated by
\begin{equation}
  v'_s = min(V,v_s) = V min\big( 1 , \frac{v_s}{V} \big)\ .
  \label{v'}
\end{equation}
It can in no way be assumed that the balls ${\cal B}_s$ are all small
and that $v'_s = v_s $.
In contrast to Voronoi tessellations, a wide dispersion of size of balls can occur, and the case when
some balls reach the size of the volume must be taken into account.

The cost of the first step of the algorithm thus reads
\begin{equation}
    n_{step~1} = N_v \sum_s min\big( 1 , \frac{v_s}{V} \big) \ .
\end{equation}

The cost of the second step depends on the number of voxels which have not been assigned to a cell in
step 1, or, in other words, which are located outside all the investigation balls.
The probability for a point $\bs{x}$ to be outside a given ball ${\cal B}_s$ is given by
\begin{equation*}
  p_s = 1 - v'_s/V \ ,
\end{equation*}
which can be approximated, using relation \ref{v'}, by
\begin{equation}
  p_s = \big( 1 - v_s/V \big)^+ \ .
\end{equation}
Under the assumption that the position and size of the balls are uncorrelated,
the probability for a given point $\bs{x}$ to be outside all the balls is approximated by
\begin{equation}
    \Pi_{s} \big( 1-\frac{v_s}{V} \big)^+ \ .
\end{equation}
The number of voxels that have to be taken into account in step 2 of the algorithm can therefore be
estimated by
\begin{equation}
  N_v \ \Pi_{s} \big( 1-\frac{v_s}{V} \big)^+ \ ,
\end{equation}
and the mean of the cost of step 2 can be estimated by
\begin{equation}
  E(n_{step~2}) = N_s N_v \ \Pi_{s} \big( 1-\frac{v_s}{V} \big)^+ \ .
\end{equation}
Finally, the estimation of the total cost of the algorithm is given by
\begin{equation}
  E(n_{step~1+2}) = N_v   \frac{ \sum_s v'_s}{V} \ + \ N_s N_v \ \Pi_{s} \big( 1-\frac{v'_s}{V} \big)
  \ ,
  \label{nstep12_jmak}
\end{equation}
with $v'_s$ given by expression \ref{v'} .

The cost of step 1 is an increasing function of $t_0$ which varies from $0$ when $t_0=min_s( t_s)$
to a maximum value
when the radial growth of the first site to appear (i.e. $arg \ \big(min_s (t_s)\big)$)
exceeds the size of the domain, i.e. at a time given by
\begin{equation*}
  t_0 = min_s( t_s) + \frac{1}{2G} \sqrt{\sum_{i=1}^d L_i^2} \ ,
\end{equation*}
in the case of periodic conditions, or at a time $t_0$ verifying
\begin{equation*}
  t_0 \le min_s( t_s) + \frac{1}{G} \sqrt{\sum_{i=1}^d L_i^2} \ ,
\end{equation*}
in the case of non-periodic conditions.

The cost of step 2 is a decreasing function of $t_0$ that varies from $N_s N_v$ when $t_0=min_s (t_s)$, to
$0$ when
$t_0 = min_s( t_s) + \frac{1}{2G} \sqrt{\sum_{i=1}^d L_i^2}$ in the periodic case,
or at a time $t_0$ lower than $min_s( t_s) + \frac{1}{G} \sqrt{\sum_{i=1}^d L_i^2}$ in the non-periodic case.
Thus the optimal value for $t_0$, i.e. the value of $t_0$ that minimises $ n_{step~1+2}$ given by relation \ref{nstep12_jmak},
must be searched between these two extreme values, e.g. with a dichotomy method,
for a negligible computational cost.

\begin{figure}[htp!]
  
  \begin{center}
    \includegraphics[angle=-90,width=7.2cm]{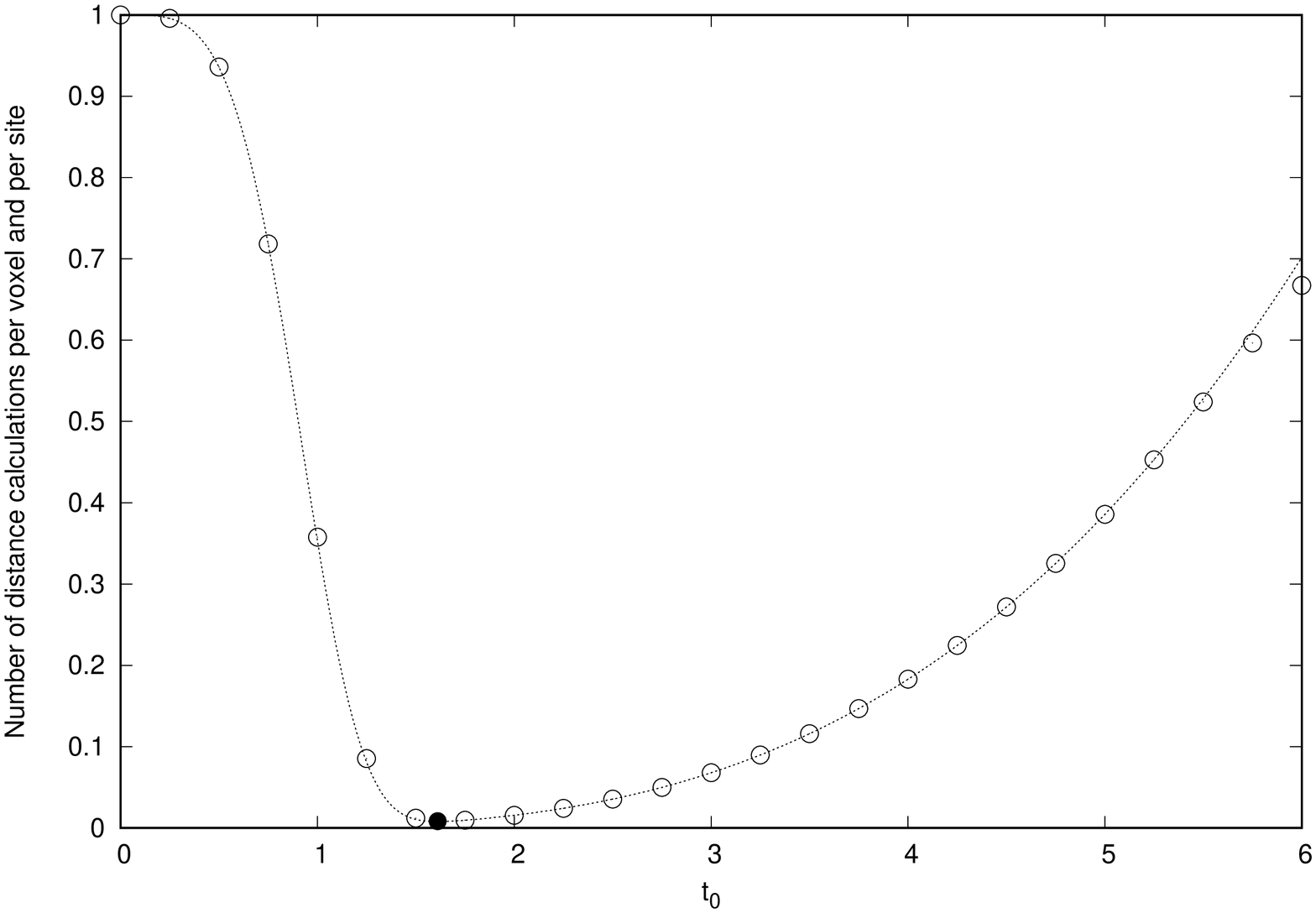}
    \includegraphics[angle=-90,width=7.2cm]{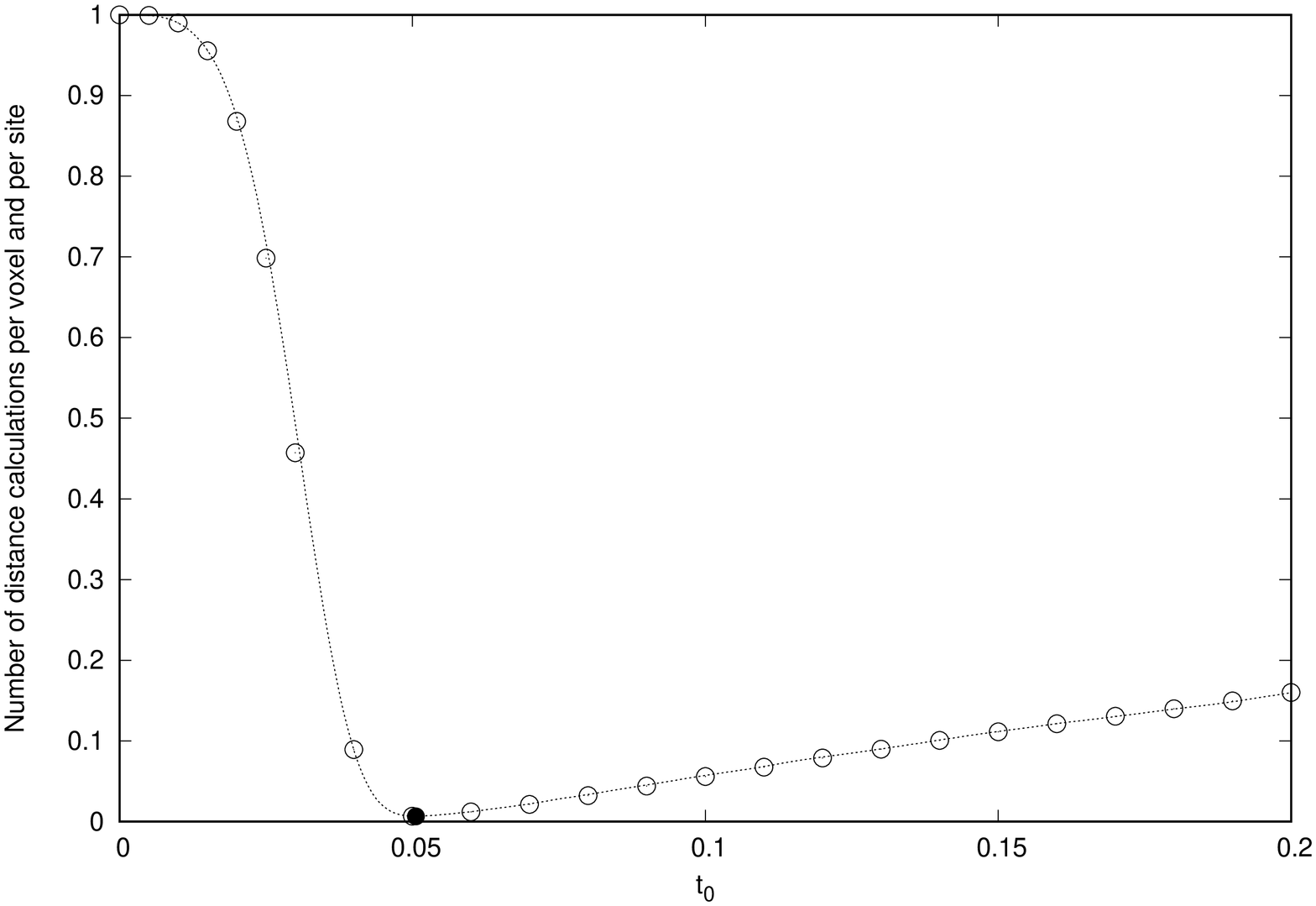}
  \end{center}
  
  \caption{
    Generation of an image of $200 \times 200 \times 200$ voxels
    of a Johnson-Mehl tessellation of  $10000$ sites. 
    Number of calculations of distances between voxels and sites in Algorithm
    \ref{faster_jmak_algo}.
    Model given by relation \ref{nstep12_jmak} (continuous line) with varying values of $t_0$.
    Measure of the numbers of
    distance calculations actually carried out in a given realisation with the algorithm (empty circles).
    Left: growth rate $G=0.1$, the optimal choice found with the algorithm is $t_0=1.57$,
    right: growth rate $G=10$,  the optimal choice found with the algorithm is $t_0=0.051$.
    Optimal value of $t_0$ estimated following the method described in section
    \ref{section_optimal_choice_Johnson_Mehl}
    (black circle).
  }
  \label{jmak_example}
\end{figure}

Figure \ref{jmak_example} illustrates the computational cost of the algorithm in a three-dimensional example.  An image
of $200 \times 200 \times 200$ voxels of a Johnson-Mehl tessellation with $10 \, 000$ sites is created.  The slight
discrepancy observed for high values of $t_0$ between the model and the data, for $G=0.1$, is most likely due to the
approximation made by the model on the volume of the investigation balls when they cross the domain boundaries.
This is consistent with the fact that the volume of the balls are overestimated by expression \ref{v'}.
Nevertheless, this does not affect the estimation of the optimal $t_0$ which is very satisfactory.

\paragraph{Remark}
Before applying the algorithm, it can be useful to rearrange the list of the sites.
Indeed, in some cases, some sites may result in the creation of no cell because they
have been reached by the crystal growth of another site before they started to grow themselves.
This happens typically when the growth rate $G$ is large.
To reduce the computation time, it is advantageous to discard these ineffective sites from
the set in a preliminary step.

\begin{algorithm}
  \SetAlgoLined
  \DontPrintSemicolon
  \SetKwInOut{Input}{input}\SetKwInOut{Output}{output}
  \SetKwInOut{Temporary}{temporary}
  
  \Input{
    \begin{itemize}
    \item a table of $N_s$ sites of positions and times $\{ (\bs{x}_s,t_s) \}$ in space domain
      $\Omega=[0,L_1) \times [0,L_2), ... \times [0,L_d)$ and time interval $[0,T)$,
              \item parameter $G$: ``crystal growth rate'' 
    \end{itemize}
    }
  \Output{
    \begin{itemize}
      \item {an image ${\cal I}(\bs{x})$ of $\bs{n}=(n_1,n_2, ...,n_d)$ voxels describing the Johnson-Mehl tessellation
        in space domain $\Omega$}
    \end{itemize}
  }
  \Temporary{
    \begin{itemize}
      \item { an image ${\cal D}$ of same dimension as $\cal I$
        }
    \end{itemize}
  }
  \BlankLine
  Initialisation: \;
  $t_0$ is estimated using the method described in section \ref{section_optimal_choice_Johnson_Mehl}
  
  \ForAll{ $\bs{k} \in \{1,...,n_1\}\times\{1,...,n_2\}\times\{1,...,n_3\}$}{
    ${\cal D}(\bs{k}) = \infty$ \;
  } 
  \BlankLine
  Step 1: \;
  \ForAll{ $s = 1, 2, ..., N_s$ }{
    \ForAll{ $\bs{k} \ \ | \  \ \bs{x(k)} \in
      (\text{Ball of center} \ x_s \ \text{and  radius}  \ r_s=G (t_0-t_s)^+) $} {
          \If{ $d(\bs{x(k)},\bs{x_s})/G + t_s < {\cal D}(\bs{k})$ }
             {
               ${\cal I}(\bs{k}) =  s$ \;
               ${\cal D}(\bs{k}) =  d(\bs{x(k)},\bs{x_s})/G+t_s$ \;
             }
        }         
  }
  \BlankLine
  Step 2: \;
  \ForAll{$\bs{k} \in \{1,...,n_1\}\times\{1,...,n_2\}\times ... \times\{1,...,n_d\}$}{
    \If{${\cal D}(\bs{k}) = \infty$}{
      \ForAll{$s = 1, 2, ..., N_s$}{
        \If{ $d(\bs{x(k)},\bs{x_s})/G + t_s < {\cal D}(\bs{k})$}{
          ${\cal I}(\bs{k}) =  s$ \;
          ${\cal D}(\bs{k}) =  d(\bs{x(k)},\bs{x_s})/G+t_s$ \;
        }
      }
    }
  }
  \caption{Faster algorithm to create an image of a Johnson-Mehl tessellation}
  \label{faster_jmak_algo}
\end{algorithm}

\subsection{An accelerated Laguerre tessellation algorithm}

The algorithm proposed for Laguerre tessellations is almost identical to the one of Jonhnson-Mehl
tessellations, except that the investigations balls, corresponding to the crystal growth at a given
time $t_0$, have radii given by
\begin{equation}
  r_s = \sqrt{G (t_0-t_s)^+} 
  \label{rs Laguerre}
\end{equation}
and that the function to evaluate the proximity of a point $\bs{x}$ to a site $\bs{x_s}$,
is given by relation \ref{LAGUERRE1}.
Moreover, the optimal choice of $t_0$ must be searched between
$min_s (t_s)$ and
$min_s( t_s) + \frac{1}{4G} {\sum_{i=1}^d L_i^2}$ in the periodic case,
$min_s( t_s) + \frac{1}{G} {\sum_{i=1}^d L_i^2}$ in the non-periodic case.

\begin{figure}[htp!]

\begin{center}
  \includegraphics[angle=-90,width=7.2cm]{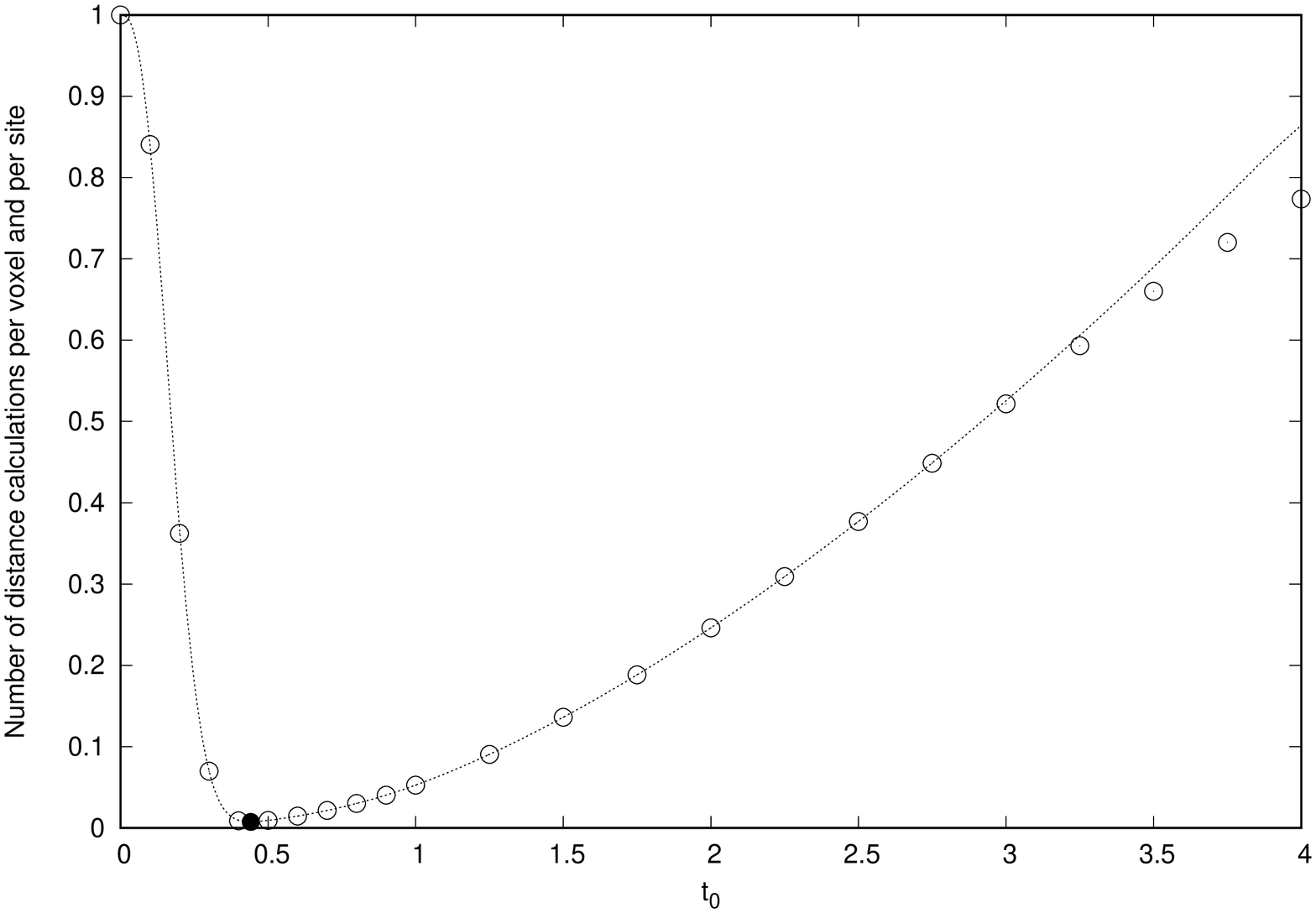}
  \includegraphics[angle=-90,width=7.2cm]{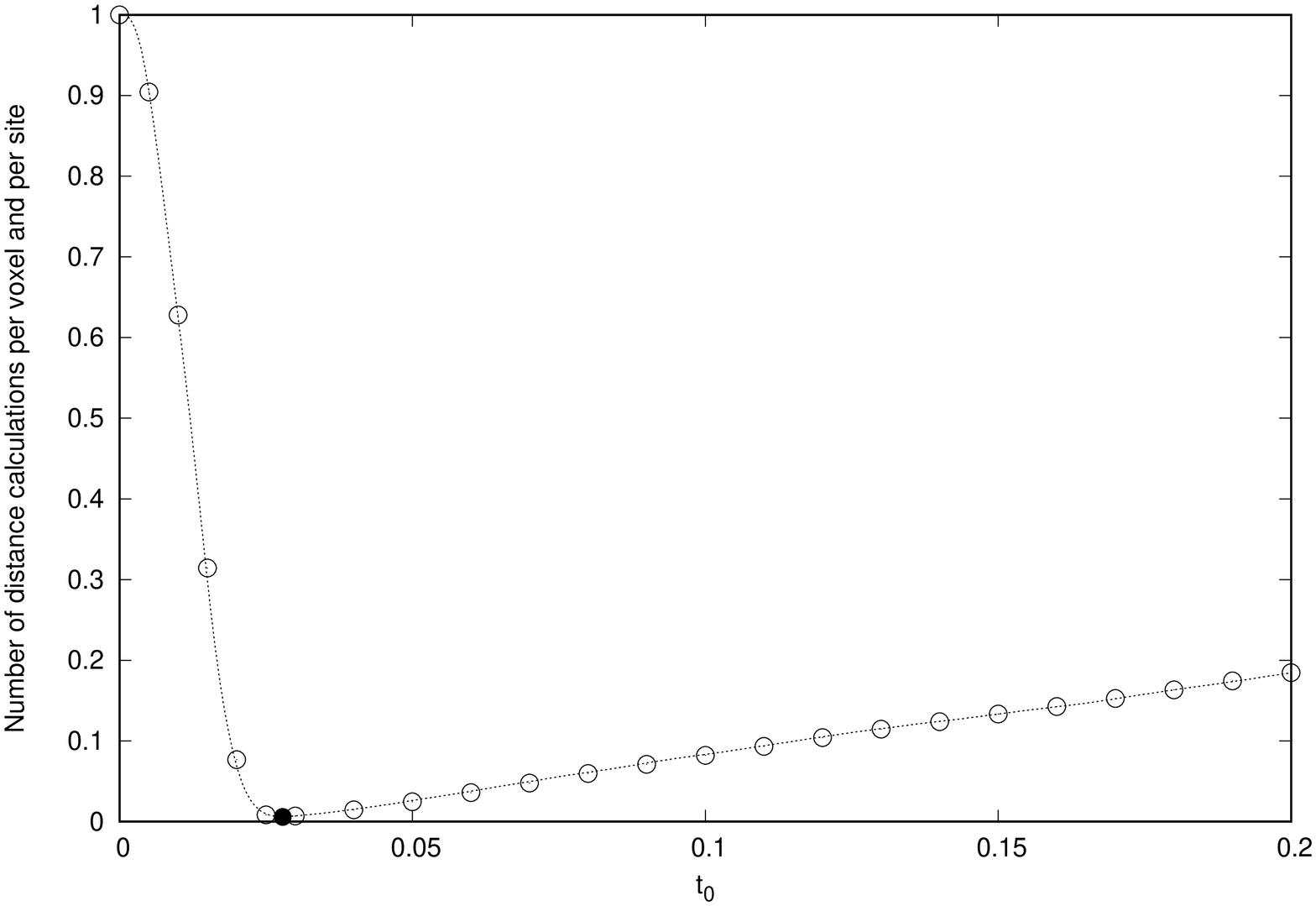}
\end{center}

\caption{ Generation of an image of $200 \times 200 \times 200$ voxels
  of a Laguerre tessellation of $10000$ sites.
  Number of calculations of distances between voxels and sites in the algorithm.
  Model giving by relation \ref{nstep12_jmak} and \ref{rs Laguerre} (dashed line) with varying values of $t_0$.
  Measure of the numbers of
  distance calculations actually performed during the execution of the algorithm (empty circles).
  Left: growth rate $G=0.1$, the optimal choice of $t_0$ is $t_0=0.426$ (black circle),
  right: growth rate $G=10$,  the optimal choice of $t_0$ is $t_0=0.028$ (black circle).
  }
\label{laguerre_example}
\end{figure}

An illustration of the behaviour of the algorithm is given in Figure \ref{laguerre_example}.
Simarly to the Johnson-Mehl algorithm, the optimal value of $t_0$ is well estimated by the model,
although a slight discrepancy (here for $G=0.1$) between the model and the data
for high values of $t_0$, which is likely due to the approximation made in the model in relation \ref{v'}.

\begin{figure}[htp!]

\begin{center}
  \includegraphics[angle=-90,width=12cm]{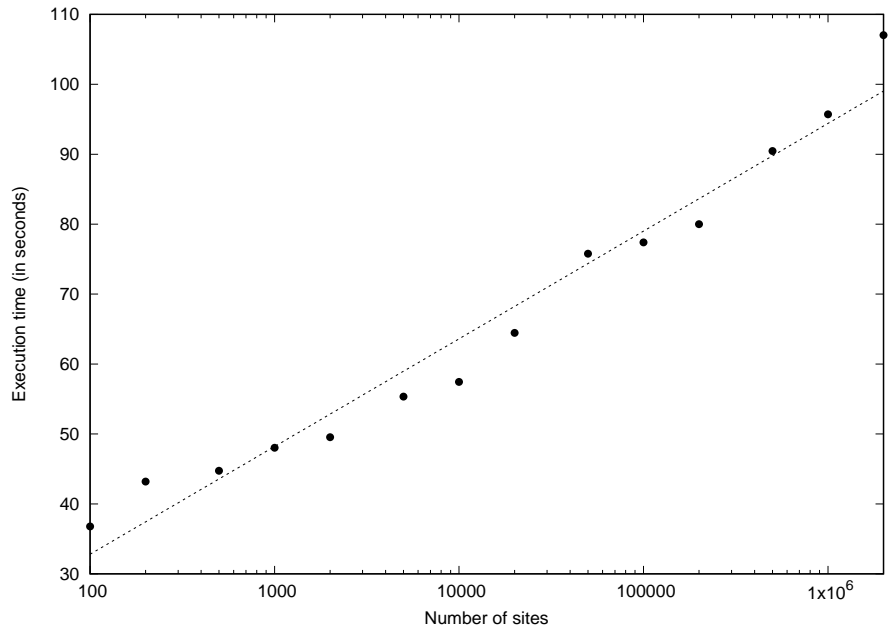}
\end{center}

\caption{ Total execution times $t_e$ of the accelerated algorithm for the creation of an image of 
  $1000 \times 1000 \times 1000$ voxels 
  of a Laguerre tessellation (radial growth $G=0.01$) for different numbers of sites $N_s$
  on a 8-core processor Xeon Silver 4208 (2.10 GHz)
  Dashed line: least square regression on the data $\big( \ln N_s , t_e \big)$.
  }
\label{laguerre}
\end{figure}

An illustration of the performance of the algorithm as a function of the number of sites is presented in Figure
\ref{laguerre}.
  
\section{Comparison with the Neper code}
\label{section comparison}

The performance of the algorithm proposed in this article is compared to that of the general code Neper (version 4.2.0.)
which is widely used in the mechanics community
involved in numerical simulations of crystalline materials.
Neper is a versatile code developed by Romain Quey (\cite{Quey_2011}) which enables to
generate artificial polycrystalline microstructure of Voronoi type,
and to create a mesh, typically to apply a finite element calculation, or to create an image of the tessellation,
for example to visualise it or to apply a calculation using an FFT-based homogenisation method.

Neper, which offers many more general features than the algorithms proposed here,
can be applied to the creation of rasterised images of tessellations.
Schematically, in that case, Neper proceeds in two successive steps.
In a first step, starting from the position of the sites of the Voronoi tessellation,
the geometry of the tessellation is determined, i.e. the vertices, edges and facets of the polyhedral cells are
determined and stored, using a so-called ``cell-based algorithm''.
In a second step, the image is created by a rasterisation process in which the cell to which
each voxel of the image belongs is determined (\cite{Quey_discussion_2021}).

\begin{figure}[htp!]

\begin{center}
  \includegraphics[angle=-90,width=12cm]{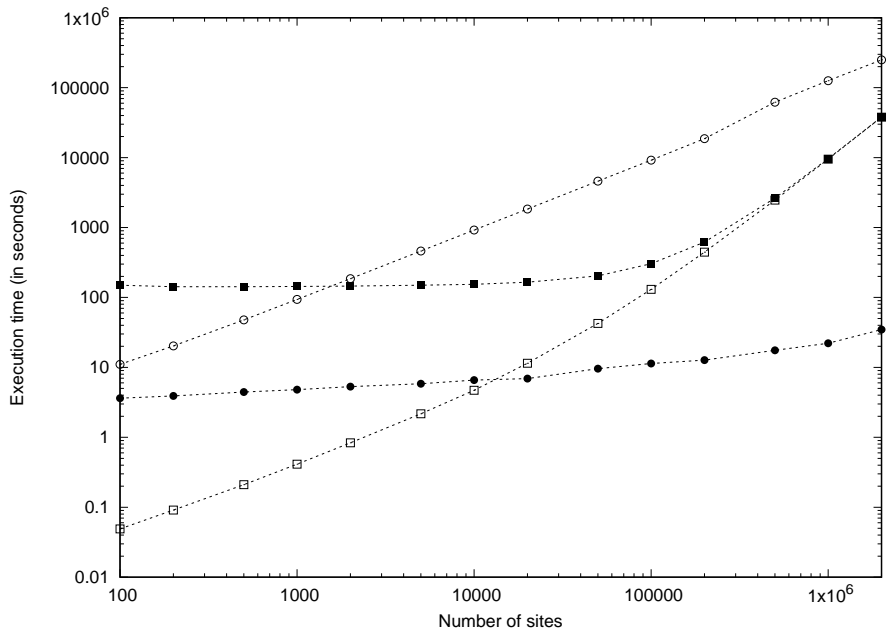}
\end{center}

\caption{ Code performance comparison on a 8-core processor Xeon Silver 4208 (2.10 GHz).
  Generation of an image of $500 \times 500 \times 500$ voxels
  of a Voronoi tessellation with different numbers of sites.
  Total execution time in seconds for the brute force algorithm (empty circles), for the accelerated algorithm
  (black circles), for Neper code (black squares), for Neper code without image rasterisation (empty squares).
  }
\label{comparison Neper}
\end{figure}

The brute force algorithm described in section \ref{brute force algo}, the algorithm outlined in section
\ref{accelerated algo} and the code Neper were applied to create a three-dimensional image of a Voronoi tessellation
containing a number of sites varying from $100$ to $2 \, 000 \, 000$. In each case, the same set of sites was used by
the three algorithms.  The dimension of the images has been fixed to $500 \times 500 \times 500$, which is a realistic
size for common calculations using an FFT-based homogenisation method.  Figure \ref{comparison Neper} shows the
execution times of the different algorithms, all three parallelised with OpenMP, on a 8-core processor Xeon Silver
4208 (2.10 GHz).

The code Neper was applied in two configurations depending on whether only the step of
creating the geometrical description of the cells was carried out
(empty squares on Figure \ref{comparison Neper}), 
or whether this step was followed by the creation of an image by ``rasterisation'' (black squares on Figure \ref{comparison Neper}).
It can be seen that the time to create the image from the geometrical description of the cells
remains approximately constant at around 140 seconds for the different site numbers,
but that the time taken to create the geometrical description increases with the number of sites:
beyond about $10 \, 000$ sites,
this time alone exceeds the time it takes to execute the accelerated Algorithm \ref{accelerated algo}.

Unsurprisingly, the brute force algorithm has low performance and the time it takes to run is proportional
to the number of sites. Nevertheless , for a low number of sites (less than $1000$), and in this specific case,
it can compete with Neper.

The accelerated code is better in all cases and,
as expected, takes an execution time of the order of the logarithm of the number of sites,
which increases its relative performance as the number of sites increases.
It took less than $30$ seconds to create an image of $2 \, 000 \, 000$ cells.

\section{Conclusion}

In this paper, an algorithm has been presented for creating digital images of Voronoi, Johnson-Mehl and
Laguerre tessellations of punctual sites assumed to be randomly distributed in the domain considered.
The principle of the algorithm is to assign the voxels belonging to balls centred around the sites of the
tessellation, to the cell corresponding to the closest of these sites.
The choice of the radii of these investigation balls plays a predominent role on the performance of the algorithm.
An analytical development
enables to find a close to optimal choice, whose validity has been verified by numerical tests.

Incidentally, a relevant definition of the distance is introduced for the case of a periodic microstructure
and a simple expression is given to calculate it.

\section*{Available code}
A beta version of the code applying the algorithms presented in this article is freely available on the site: 

\url{https://lma-software-craft.cnrs.fr/voronoi-johnson-mehl-and-laguerre-tessellations} .

\section*{Acknowledgments}
The author thanks the ``Institut de Radioprotection et de Sûreté Nucléaire'', and especially Dr Pierre-Guy Vincent,
who has supported the project for which the method proposed in this article has been developed.

The author also thanks Romain Quey for his interest in this study and for his pertinent advice.
\appendix
\section{L-periodic distance}
\label{L-distance appendix}

\subsection{Periodicity}

In the Euclidean space of dimension $d$, a given property $f$ defined at each point $\bs{x}$ is periodic of
periods $\bs{L}_i$, with $i=1, 2, ..., d$, when
\begin{equation}
  f( \bs{x} ) = f(\bs{x} + \sum_{i=1}^d k_i \bs{L_i}) \quad \forall k_i \in \mathbb{Z} \ ,
\end{equation}
where $\bs{L}_i$ are given $d$-dimensional vectors supposed to be linearly independent of each other.

The property $f$ is completely determined by its definition in  the unit cell $\cal U$
\begin{equation*}
  {\cal U} = \big\{ \bs{x}=\sum_{i=1}^{d} \alpha_i \bs{L_i} \quad \
  | \  \quad \alpha_i \in [O,1) \ \ \big\} \ ,
\end{equation*}
or in any translation of ${\cal U}$ by a vector $\bs{o}$:
\begin{equation}
  {\cal L} = \big\{ \bs{x}=\bs{o} + \sum_{i=1}^{d} \alpha_i \bs{L_i}
  \quad \ | \  \quad \alpha_i \in [O,1) \ \ \big\} \ .
      \label{def_unit_cell}
\end{equation}

\subsection{Definition of the L-periodic distance}
\begin{figure}[htp!]
  \begin{center}
    \includegraphics[width=12cm]{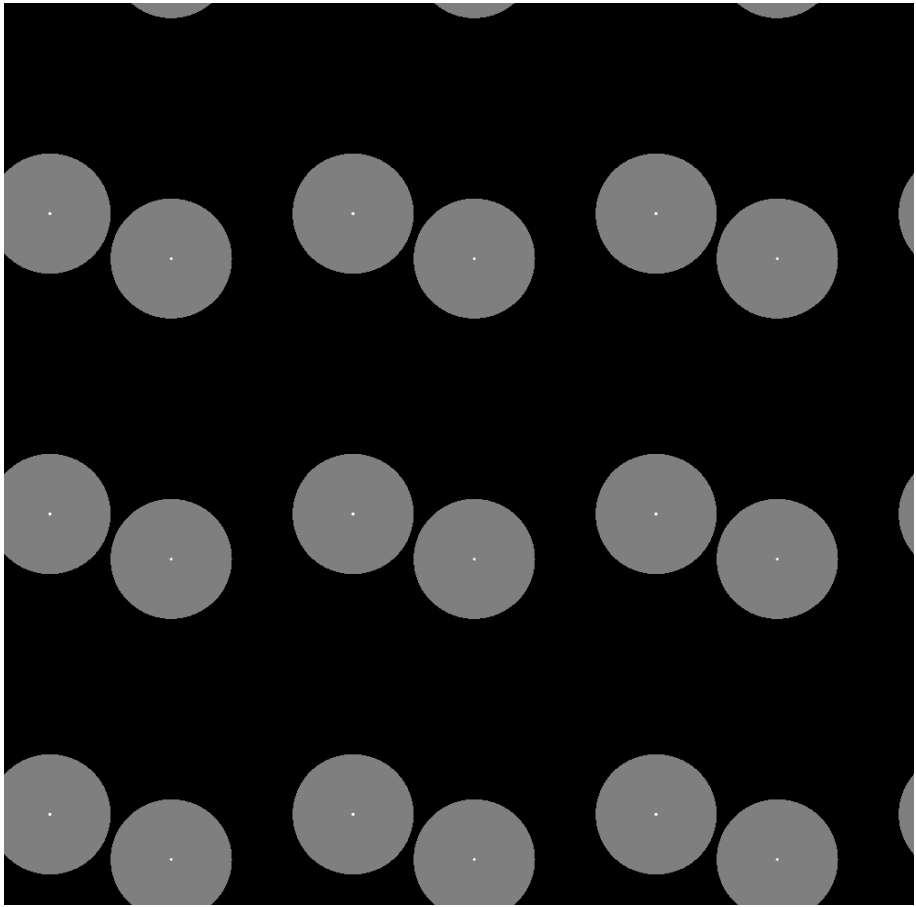}
  \end{center}  
  \caption{
    Two-dimensional microstructure consisting in two similar discs (in gray) periodically
    repeated in a matric (black).
    The centres of the discs are represented in small white circles.
    This figure repoduces $3 \times 3$ periods of the pattern.
  }
  \label{two discs a}
\end{figure}

\begin{figure}[htp!]
  \begin{center}
    \includegraphics[width=7.2cm]{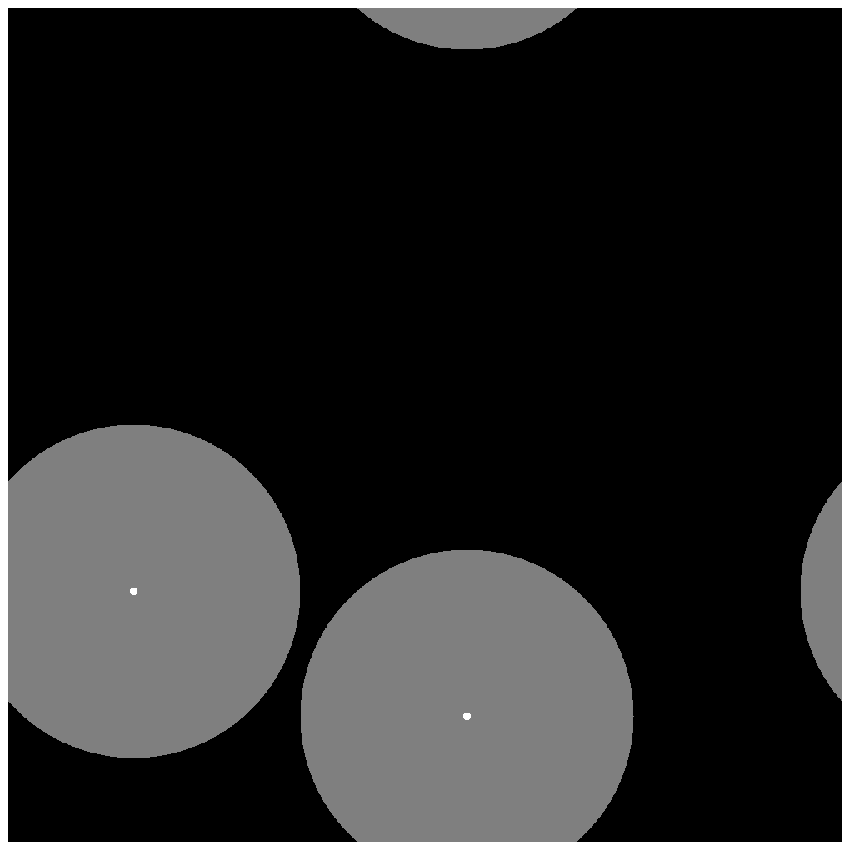}
    \hspace{0.3cm}
    \includegraphics[width=7.2cm]{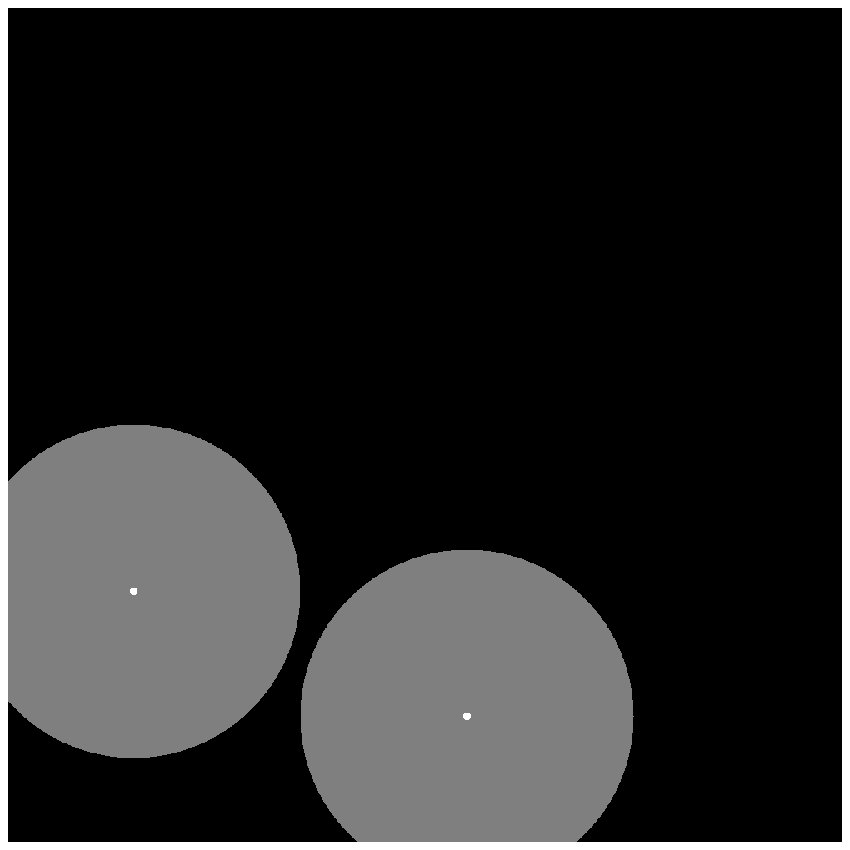}
  \end{center}  
  \caption{
    Left: unit cell of the two-dimensional microstructure \ref{two discs a}.
    Right: image of the same discs {\it without} periodic conditions.
  }
  \label{two discs b}
\end{figure}

To introduce the definition of the L-periodic distance, the simple example of a 
two-dimensional microstructure containing two circular inclusions periodically repeated in a matrix
is considered.
An illustration
of the resulting microstructure is presented in Figure \ref{two discs a} where a period of the pattern is repeated
three times in each direction.
The left picture of Figure \ref{two discs b} represents a unit cell of the microstructure.
By comparison with the microstructure with the same two discs but without periodic conditions
(Figure \ref{two discs b} right), 
it can be seen that the parts of the discs that leave the image by one edge (here on the left and bottom edge) reappear on the opposite edge.

Thus, building an image of a unit cell under non-periodic conditions, consists in determining, for each of its voxels, if
it belongs to a ball (or disc in 2D) by comparing the Euclidean distance between the voxel and the centre of the ball to
the radius of the latter. In the case of periodic conditions, one has to compare the distance between the location
$\bs{x}$ of each voxel and every periodically repeated centres of the balls, to the radius.  In other words, one has to
evaluate
\begin{equation}
      d'({\bs{x},\bs{x_s}}) \ 
      = \ min_{\bs{k}\in {\mathbb{Z}}^d}
      \big( d_E( \bs{x}, \bs{x_s}+\sum_{i=1}^d {k_i} \bs{L}_i )  \big)
      \ ,
\end{equation}
where $\bs{x}$ denotes the position of the voxel and $\bs{x_s}$ the one of the site or centre,
where $\bs{k}$ is a d-dimensional vector of integer components $k_i$,
and where $d_E$ denotes the usual Euclidean distance:
\begin{equation}
  d_{E}({\bs{x},\bs{y}}) \ = \ \sqrt{ \sum_{i=1}^d (x_i-y_i)^2} \ ,
  \label{de}
\end{equation}
where $x_i$ and $y_i$ are the components of vectors $\bs{x}$ and $\bs{y}$ in an orthonormal basis.

In the particular case where
the sites are inside the unit cell and that the ball diameter is smaller than the dimensions of the unit cell,
it is sufficient to estimate the distances between the voxels of the image of the unit cell and the ball centres
which has been repeated by one period forward and backward in every directions
(\cite{Fritzen_2009}, \cite{Bargmann_2018} or \cite{Yan_2011}), and one has
\begin{equation}
      d'({\bs{x},\bs{x_s}}) \ 
      = \ min_{\bs{k}\in  \{-1,0,1\}^d}
      \big( d_E( \bs{x}, \bs{x_s}+\sum_{i=1}^d {k_i} \bs{L}_i )  \big)
      \label{d'}
\end{equation}
(each component $k_i$ of $\bs{k}$ only takes the value $-1$, $0$ or $1$).
 Thus, to evaluate expression \ref{d'}, one has to calculate
 9 Euclidean distances in a two-dimensional problem, and 27 Euclidean distances in a three-dimensional problem.


However, this approach does not work in the more general case where the points whose distance is to be calculated can be
located outside the unit cell.
This typically occurs in the case of a microstructure that contains inclusions whose size exceeds one of the
dimensions of the unit cell (for example long fibers in a periodic microstructure).

Therefore, we define the L-periodic distance between two points of coordinates $\bs{x}$ and $\bs{y}$ of the
Euclidean space - which can be located outside the unit cell - where the property
considered is supposed to be periodic, with periods $\bs{L_i}$, with $i=1,2,...,d$, by
\begin{equation}
    d_{L}({\bs{x},\bs{y}}) \ 
    \defeq \ min_{\bs{k}\in {\mathbb{Z}}^d,\bs{k'}\in {\mathbb{Z}}^d }
    \big( d_E( \bs{x} + \sum_{i=1}^d k_i \bs{L}_i , \bs{y}+\sum_{i=1}^d {k'_i} \bs{L}_i )  \big)
    \ .
  \label{L-distance_def}
\end{equation}

\subsection{Calculation of the L-periodic distance}

In this paragraph, the expression of the L-periodic distance is developed to obtain a formulation
convenient to use in practice.

It is straightforward that relation \ref{L-distance_def} simplifies into 
\begin{equation*}
    d_{{L}}({\bs{x},\bs{y}})
    = min_{\bs{k}\in {\mathbb{Z}}^d}
    \Big( d_E \big( \bs{x} , \bs{y}+\sum_{i=1}^d {k_i} \bs{L}_i \big)  \Big) \ .
\end{equation*}
In the case when the vectors $\bs{L_i}$
are all orthogonal to each other, they can be noted
\begin{equation*}
  \bs{L}_i =  L_i \bs{e}_i \ ,
\end{equation*}
where $L_i$ are positive scalars and $\bs{e}_i$ are the unit vectors of an orthonormal basis.
Hence, $d_L$ can be written
\begin{equation*}
  d_{{L}}^2({\bs{x},\bs{y}}) =
  \sum_{i=1}^d 
    min_{k_i \in \mathbb{Z}} \big(
     y_i - x_i + k_i L_i
    \big)^2
    \ ,
\end{equation*}
where $x_i$ and $y_i$ are the components of $\bs{x}$ and $\bs{y}$ in the basis $(\bs{e_i})$.

With no real difficulties, it can be established that
$$
min_{k_i \in \mathbb{Z}}\big( ( y_i-x_i + k_iL_i)^2 \big) =
\bigg( y_i-x_i - \text{N}\Big( \frac{y_i-x_i}{L} \Big) L \bigg)^2
\ ,
$$
where $N()$ is the ``nearest integer function'' defined as
$$\text{N}(x)=\lfloor x + \frac{1}{2} \rfloor$$
where  $\lfloor . \rfloor$ denotes the ``floor function'', that gives the greatest integer
value lower or equal than its real number argument.
The nearest integer function can be calculated using the function {\tt rint()} of the C mathematical library.

Finally, the L-periodic distance $d_L$ can be written 
\begin{equation}
  d_{{L}}({\bs{x},\bs{y}}) =
  \sqrt{
  \sum_{i=1}^d
    \bigg(
    y_i - x_i - \text{N} \Big( \frac{y_i -  x_i}{L_i}  \Big) L_i
    \bigg)^2
    } \ .
    \label{dL}
\end{equation}

The numerical cost of the calculation of the L-periodic distance using expression \ref{dL} is only slightly
higher than the one of the Euclidean distance with expression \ref{de}. In addition to being an exact method,
it is very simple to implement and its numerical cost is much lower than evaluating the minimum of 27
(respectively 9) Euclidean distances in 3D (respectively 2D) problems, as proposed by \cite{Fritzen_2009}
and \cite{Bargmann_2018}.

\section{Alternative definition of Laguerre tessellations}
\label{Laguerre_appendix}
As already presented in section \ref{Laguerre_section},
a Laguerre tessellation associated with the set of sites
$  {\cal S} = \{ (\bs{x_s}, r_s) \}_{s=1,2,...,N_s} $
where $\bs{x_s}$ is the position of site $s$ and $r_s$ is an associated positive scalar,
is a division of the domain ${\Omega}$ of the Euclidean space into $N_s$ cells defined by
\begin{equation}
  C_s = \{
  \bs{x} \in {\Omega} \ | \
  p_d(\bs{x},s)
  \le
  p_d(\bs{x},s') , \forall s' \in {\cal S}
    \}
\end{equation}
with the so-called power-distance $p_d$ being defined by
\begin{equation}
  p_d(\bs{x},s) = d^2(\bs{x},\bs{x_s}) -r_s^2 .
\end{equation}

For a given growth constant $G$, one can define  a time $t_s$ for each site $s$ as
$$ t_s = - r_s^2/G$$
and thus each cell $C_s$ can be explicited as
\begin{equation}
  C_s = \{
  \bs{x} \in {\Omega} \ | \
    d^2(\bs{x},\bs{x_s}) + G t_s \le d^2(\bs{x}, \bs{x_{s'}})+ G t_{s'}, \forall s' \in {\cal S}
    \}
    \ .
\end{equation}
The time $t_s$ at which a site ``appears'', if one takes again the analogy with the crystallisation process,
does not necessarily have to be negative
if one remarks
that the partition obtained does not depend on the choice of the time reference,
i.e. the partition is unchanged if one considers time $t' = t-t_{ref}$ instead of $t$,
the power-distance being modified into
\begin{equation}
  d^2(\bs{x}, \bs{x}_s) + {G} t_s' = d^2(\bs{x}, \bs{x}_s) + {G} t_s - {G} t_{ref}
\end{equation}
because
\begin{equation}
  min_s \big( d^2(\bs{x}, \bs{x}_s) + {G} t_s' \big) =
  min_s \big( d^2(\bs{x}, \bs{x}_s) + {G} t_s \big)
  - {G} t_{ref} .
\end{equation}
The relation between the radius $r_s$ and time $t_s$ can then be expressed as
\begin{equation}
  {G} (t_s - t_{ref}) = -r_s^2
\end{equation}
with $t_{ref}$ chosen adequately.

\bibliography{Moulinec_MTC_2021}
\end{document}